\documentclass[fleqn,10pt]{wlscirep}
\usepackage[utf8]{inputenc}
\usepackage[T1]{fontenc}
\usepackage{tablefootnote}

\title{Reaction profiles for quantum chemistry-computed [3 + 2] cycloaddition reactions}

\author[1]{Thijs Stuyver}
\author[2,3,4]{Kjell Jorner}
\author[1,5]{Connor W. Coley}
\affil[1]{Department of Chemical Engineering, Massachusetts Institute of Technology, 77 Massachusetts Avenue, Cambridge, Massachusetts 02139, United States}
\affil[2]{Department of Computer Science, University of Toronto, 40 St George St, Toronto, Ontario M5S 2E4, Canada}
\affil[3]{Department of Chemistry, Chemical Physics Theory Group, 80 St. George St., University of Toronto, Ontario M5S 3H6, Canada}
\affil[4]{Department of Chemistry and Chemical Engineering, Chalmers University of Technology, Kemigården 4, SE-41258, Gothenburg, Sweden}
\affil[5]{Department of Electrical Engineering and Computer Science, Massachusetts Institute of Technology, 77 Massachusetts Avenue, Cambridge, Massachusetts 02139, United States}

\affil[*]{corresponding author(s): Connor Coley (ccoley@mit.edu)}

\begin{abstract} 
Bio-orthogonal click chemistry based on [3 + 2] dipolar cycloadditions has had a profound impact on the field of biochemistry and significant effort has been devoted to identify promising new candidate reactions for this purpose. To gauge whether a prospective reaction could be a suitable bio-orthogonal click reaction, information about both on- and off-target activation and reaction energies is highly valuable. Here, we use an automated workflow, based on the autodE program, to compute over 5000 reaction profiles for [3 + 2] cycloadditions involving both synthetic dipolarophiles and a set of biologically-inspired structural motifs. Based on a succinct benchmarking study, the B3LYP-D3(BJ)/def2-TZVP//B3LYP-D3(BJ)/def2-SVP level of theory was selected for the DFT calculations, and standard conditions and an (aqueous) SMD model were imposed to mimic physiological conditions. We believe that this data, as well as the presented workflow for high-throughput reaction profile computation, will be useful to screen for new bio-orthogonal reactions, as well as for the development of novel machine learning models for the prediction of chemical reactivity more broadly.
\end{abstract}
\begin{document}

\flushbottom
\maketitle

\thispagestyle{empty}

\section*{Background \& Summary} 

Bio-orthogonal click reactions provide a straightforward method to tag and image certain biomolecules, enabling the study of their function \emph{in vivo}. \cite{saxon2000cell,agard2004strain,sletten2009bioorthogonal} Besides realtime imaging, bio-orthogonal click reactions have also been considered for advanced applications in medicine/healthcare, e.g., targeted cancer treatments and \emph{in situ} drug assembly.\cite{devaraj2018future, agarwal2015site} New candidate reactions -- especially mutually orthogonal ones -- are highly sought after. \cite{kim2015bioorthogonal, sletten2011bioorthogonal, nguyen2020developing}

A wide variety of criteria need to be fulfilled for a reaction to be considered suitable for bio-orthogonal click chemistry: (a) the reactants should be stable and the reaction selective, i.e., reactant degradation due to biological conditions (pH, solvent, etc.) should be limited and side reactions with biomolecules should not occur at appreciable rates; (b) the selective reaction should be sufficiently fast at room/body temperature so that ligation can take place before clearance of the reactants from the studied organism; and (c) it should be possible to incorporate the reactants into biomolecules via some form of metabolic or protein engineering. Ideally the compounds should also be small enough to not disturb the native behavior of the biological system under investigation. \cite{jewett2010cu} As such, data regarding thermodynamic stability as well as on- and off-target kinetics could facilitate a principled screening campaign to discover novel candidate bio-orthogonal click reactions.

\begin{figure}
\centering
\includegraphics[scale=0.36]{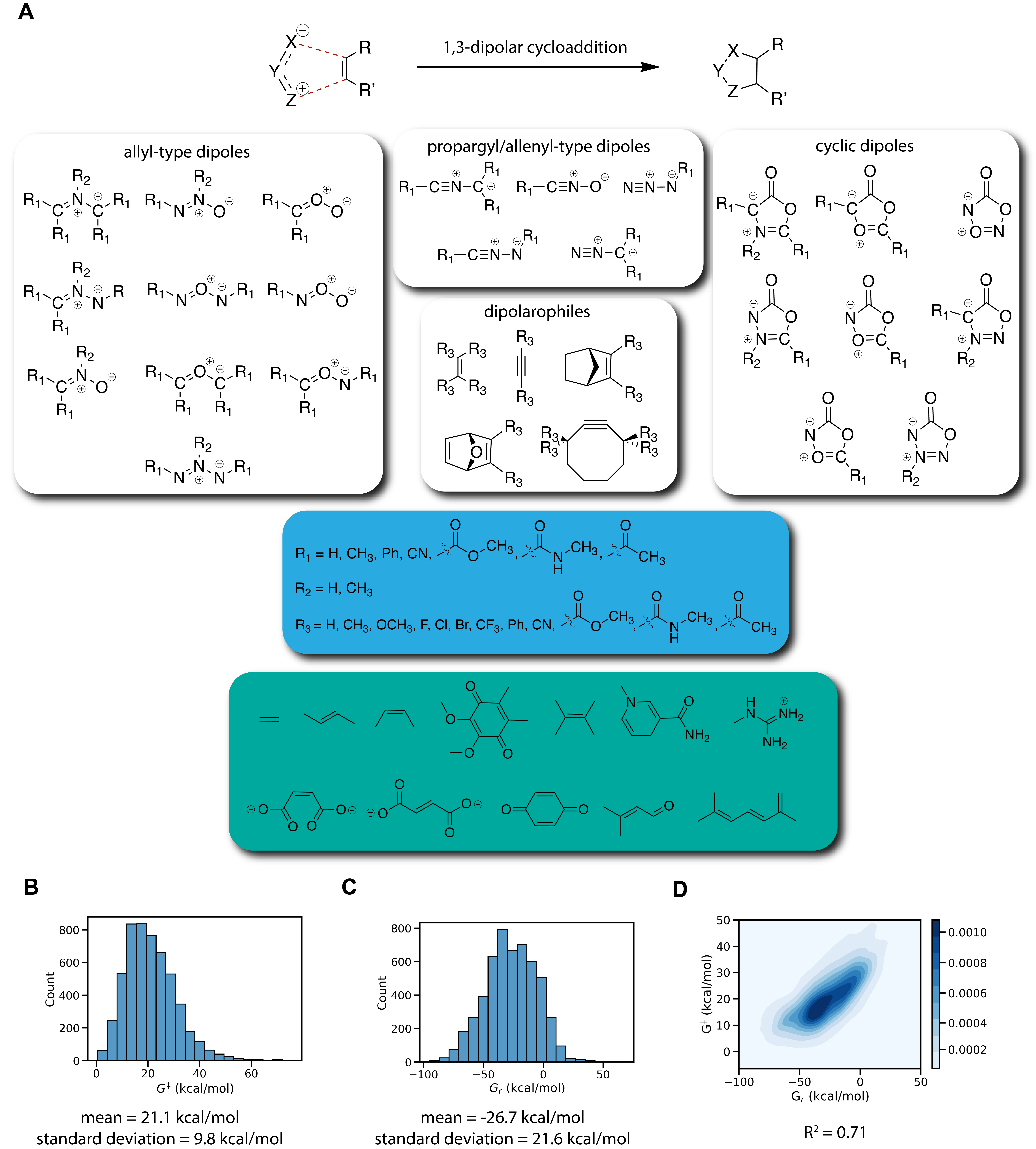}
\caption{(A) Schematic overview of the search space for the considered dipoles and dipolarophiles. In the green box, the biologically relevant motifs considered in this study are presented. These motifs are inspired respectively by fatty acids and prostaglandins, \cite{bergstrom1968prostaglandins} fumaric and maleic acid, \cite{krebs1938formation} ubiquinone or co-enzyme Q (a common coenzyme family that is ubiquitous in animals and most bacteria), \cite{ernster1995biochemical} terpineols, \cite{khaleel2018alpha} NADH/NADPH (an important and common co-enzyme playing a pivotal role in the metabolism of most species), \cite{turrens1980generation} retinal (a chromophore central to visual perception), \cite{mcilwain1996introduction}, $\beta$-carotene, \cite{burton1984beta} and the arginine amino-acid and argininosuccinic acid (an important intermediate in the urea cycle). \cite{erez2011argininosuccinate} (B) Histogram representing the distribution of the computed activation energies (G$^{\ddagger}$). (C) Histogram representing the distribution of the computed reaction energies (G$_{r}$). (D) Correlation plot between activation and reaction energies.}
\label{fig:overview}
\end{figure}

Here, we present a computational dataset of hypothetical click reactions evaluated for their potential bio-orthogonality constructed via a high-throughput reaction profile evaluation. Due to the enormous success -- and computationally accessible mechanism -- of [3 + 2] dipolar cycloaddition reactions in bio-orthogonal chemistry so far, \cite{jewett2010cu} we decided to focus on this reaction class. A diverse chemical space spanning close to 5M synthetic dipole-dipolarophile pairs was defined (Figure \ref{fig:overview}A) and the reaction profiles, i.e., activation and reaction energies, were evaluated for a representative subset of over 3000 individual reactions. We also probe the inherent reactivity of various dipoles towards a variety of "native" dipolarophiles, i.e., unsaturated motifs within biomolecules that are present in varying concentrations in living organisms. To this end, more than 2500 off-target reactions involving 12 distinct "biofragments" (green box in Figure \ref{fig:overview}A), selected from the literature, were computed as well.

A limited benchmarking, with respect to both computational reference data as well as experimental data, points to B3LYP-D3(BJ)/def2-TZVP//B3LYP-D3(BJ)/def2-SVP as the level of theory striking the optimal balance between performance and cost (\emph{vide infra}). To mimic biological environments, water was taken as the solvent, which was modelled through the SMD polarizable continuum model, and standard temperature and pressure were assumed. \cite{marenich2009universal} Histograms, depicting the distribution of the activation energies ($G^{\ddagger}$) and reaction energies ($G_r$) obtained with these settings for the 5271 successfully computed reaction profiles (out of a total of 5974 considered reaction SMILES), are provided in Figure \ref{fig:overview}B-D. A significant spread in both the activation energies and the reaction energies is clearly obtained (the standard deviations amount to 9.8 and 21.6 kcal/mol respectively), indicating that chemically diverse reacting systems were sampled. Even though the dataset contains some diversity in dipole/dipolarophile scaffolds, reaction and activation energies correlate quite well ($R^2$ = 0.71; see Figure \ref{fig:overview}D) in accordance with the Bell-Evans-Polanyi principle. \cite{evans1936further}

The produced dataset should prove useful for the construction of (surrogate) machine learning (ML) models for the discovery of promising dipole-dipolarophile combinations within the defined search space. Furthermore, the provided data will be useful for the development of methodologies for reactivity prediction in general, particularly due to the inclusion of high quality 3D geometries for each species (see below).

\section*{Methods}

To construct the dataset of reaction profiles for [3 + 2] cycloaddition reactions, an automated workflow based on the Python package autodE, \cite{autodE} recently developed by the Duarte group, has been set up. autodE is a package which can be coupled to various popular electronic structure programs to automate the otherwise laborious task of computing full reaction profiles; here, we opted for Gaussian16.\cite{frisch2016gaussian} The starting point of an autodE computation can be a simple reaction SMILES string, together with a specification of the solvent, temperature and level of theory. 

\subsection*{Definition of the chemical space of dipoles and dipolarophiles}

First, a combinatorial dataset of 1,3-dipoles and dipolarophiles was generated. For the dipoles, 15 allyl and propargyl-allenyl scaffolds, \cite{breugst2020huisgen} decorated with a variety of substituents (Figure \ref{fig:overview}A), were selected. Additionally, (decorated) heterocyclic 5-membered rings based on 1,3-dipoles, e.g. sydnone- and münchnone-derivatives, were also included. \cite{breugst2020huisgen} Since dipoles used in bio-orthogonal click applications usually require an attachment site which can be linked to either the target biomolecule or the (fluorescent) probe, only substituent combinations containing at least one substituent ending with an (extendable) methyl or phenyl group were considered. 

For the dipolarophiles, 3 strained scaffolds (cyclooctyne, norbornene and norbornadiene) \cite{jewett2010cu} as well as regular ethylene and acetylene scaffolds, each with a variety of substituents, were selected (see Figure \ref{fig:overview}A). For the ethylene and acetylene scaffold, the same constraint on substituent combination diversity as for the dipoles was introduced (the remaining dipolarophile scaffolds have several alternative attachment points which can be connected to a target molecule/fluorescent probe). Additional constraints were placed on the substituent combinations for the ethylene and cyclooctyne scaffold to reign in the combinatorial explosion of possibilities. For ethylene, the total number of distinct substituent types was fixed to two at most; for cyclooctyne, the substituents on at least one of the functionalized carbon sites was forced to be the same.

In total, 3555 dipoles and 1339 dipolarophiles were generated in this manner with the help of RDKit \cite{landrum2013rdkit}, which results in a combinatorial chemical space of almost 4.8M reacting systems and -- taking regio-isomerism into account -- 9.6M+ unique reactions. From this vast chemical space, a computationally tractable subset was extracted. To ensure representation of each individual scaffold type, both dipoles and dipolarophiles were subdivided into individual buckets and random sub-samples were taken (with replacement) from each. For the dipoles, 1000 species were selected from the allyl-based scaffolds, 200 from the propargyl-allenyl scaffolds, and 300 from the heterocyclic 5-membered rings. For both the ethylene and acetylene dipolarophile scaffolds, 200 species were sampled, whereas 300 were sampled for both the norbornene and oxo-norbornadiene scaffolds. For the cyclooctyne scaffold, 500 species were selected. 

Subsequently, the sampled list of dipolarophiles was shuffled, after which the entries of the dipole and dipolarophile lists were matched based on index values. The reactant systems generated in this manner were subsequently passed to the RunReactants function from RDChiral together with a set of SMARTS templates to generate full, atom-mapped reaction SMILES (\emph{vide infra}). \cite{coley2019rdchiral} 
\\
The SMARTS templates used are shown in Table \ref{tbl:SMARTS}.

To probe potential off-target reactivity, 12 unsaturated biologically relevant motifs that are present/are produced as metabolites \emph{in vivo} were selected.\cite{bergstrom1968prostaglandins, krebs1938formation, ernster1995biochemical, khaleel2018alpha, turrens1980generation, mcilwain1996introduction, burton1984beta, erez2011argininosuccinate} To construct a dataset of reactions involving these biofragments, a new sample of 1500 dipoles was taken in the same manner as before, as well as a sample (with replacement) of 1500 biofragments, see the green box in Figure \ref{fig:overview}A for an overview of the motifs considered. Subsequently, the sampled biofragment list was again shuffled and the entries of the two lists were matched based on index value.

\subsection*{Solvent correction and thermal effects}

\begin{figure}
\centering
\includegraphics[scale=0.35]{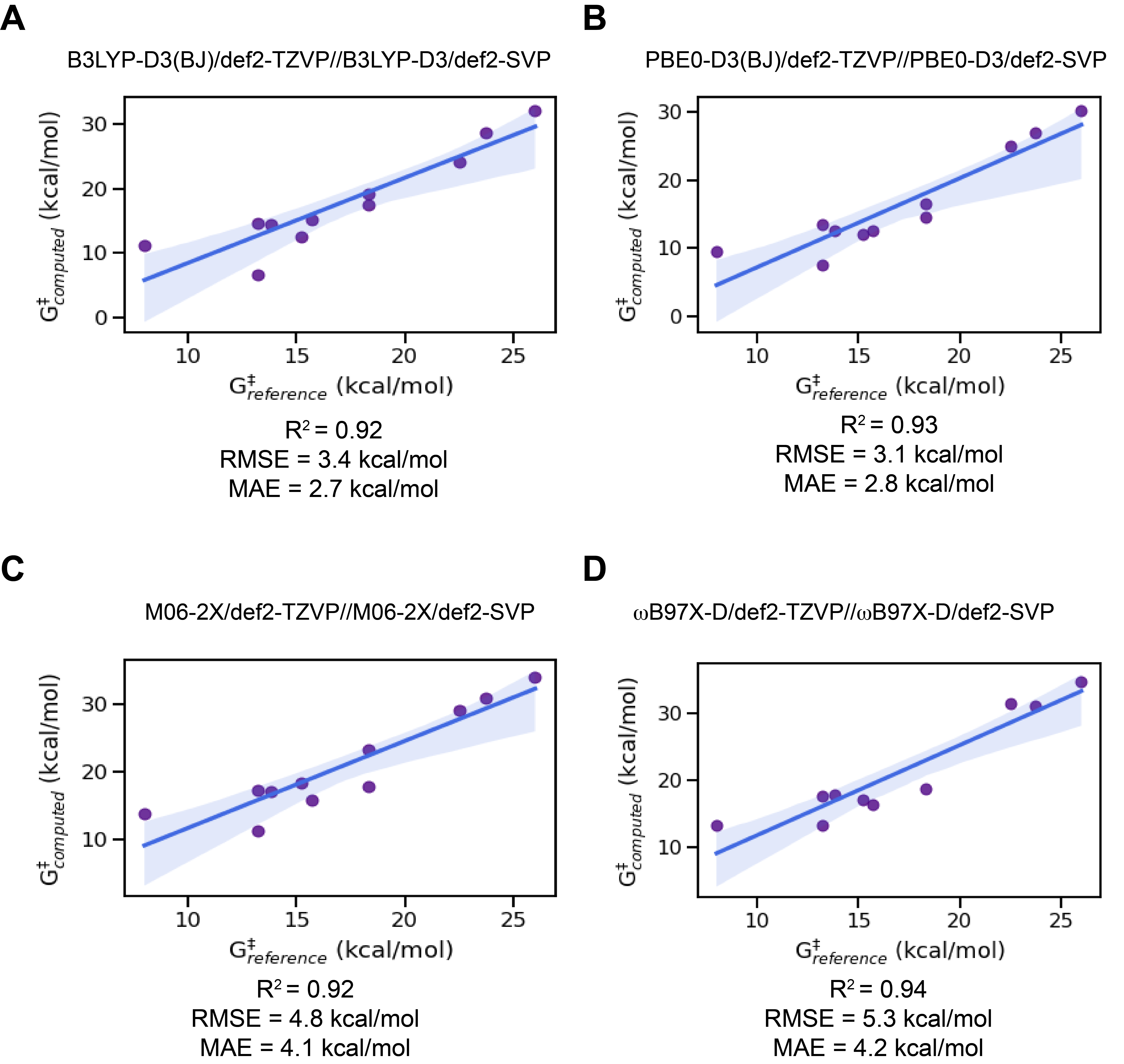}
\caption{Benchmarking functional + dispersion correction combinations against the experimental dataset extracted from the literature. The translucent bands around the regression line correspond to the 95\% confidence interval determined through bootstrap resampling.}
\label{fig:benchmarking_solvent}
\end{figure}

To mimic biological environments, we model the reaction systems with water as a solvent through the SMD polarizable continuum model. \cite{marenich2009universal} A recent computational study by Yang and co-workers investigated the role of explicit water molecules on 1,3-cycloaddition reactions using DFT-quality neural network potentials and compared to experimental data. They concluded that even in the most extreme cases considered, aqueous cycloaddition reactions tended to retain a (quasi-)concerted mechanism and most \emph{qualitative} solvation trends could already be successfully recovered through implicit modelling.\cite{yang2019role} Our own benchmarking results also corroborate the adequacy of implicit solvent modelling to recover experimental trends (Figure \ref{fig:benchmarking_solvent}). 

The choice for water as solvent has been informed by its ubiquity in biological settings, though this may not always be an appropriate approximation for the actual environment in which these click reactions will take place (e.g., in the vicinity of cell membranes or in protein condensates). It is, however, well-established that the rates of 1,3-cycloaddition reactions are fairly insensitive to solvent polarity, so that deviations from the idealized solvent environment is not an issue of major concern. \cite{huisgen1963kinetics} 

To retain the generality of our approach, we selected standard conditions for the computation of thermal corrections. The temperature of each reaction was set to room temperature (298.15K) and the concentration of each species to 1 mol/L. It should be obvious that in practice, significant deviations from standard conditions can be expected, e.g., body temperatures of most mammals hover around 310K, \cite{morrison1952weight} and, upon  entry of a (synthetic) dipole in a biological system, product concentrations will initially be negligible compared to those of the reactants, which will decrease the Gibbs free energy of reaction and hence facilitate the transformation. Nevertheless, the presented values under standard conditions are suitable reference values which can be transferred to the specific application under consideration.

\subsection*{Selection of DFT functional and basis set and validation of workflow}

The DFT level of theory was selected based on a (limited) benchmarking study. In total, 4 different functionals and dispersion correction combinations were considered (B3LYP-D3(BJ), \cite{becke1988density, lee1988development, stephens1994ab, grimme2006semiempirical} PBE0-D3(BJ), \cite{adamo1999toward} M06-2X \cite{zhao2008m06} and $\omega$B97X-D), \cite{chai2008long, chai2008systematic} in combination with the def2-SVP basis set \cite{schafer1992fully} for optimizations and frequency/thermal correction calculations, and def2-TZVP \cite{schafer1994fully} for single-point calculations. 

\begin{figure}
\centering
\includegraphics[scale=0.5]{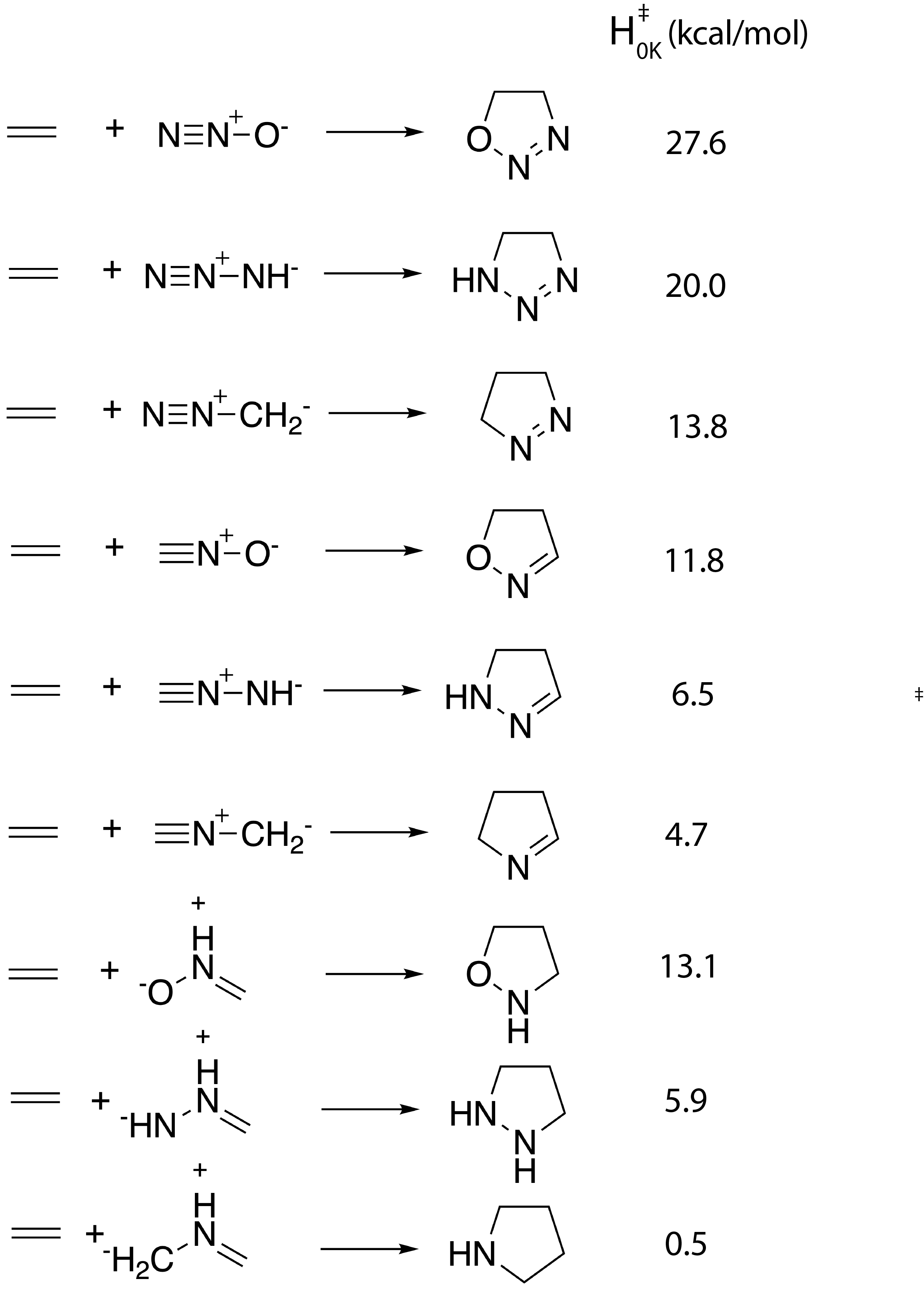}
\caption{Reference reaction data extracted from the (revised) BHPERI dataset.\cite{karton2015accurate}}
\label{fig:BHPERI}
\end{figure}

\begin{figure}
\centering
\includegraphics[scale=0.35]{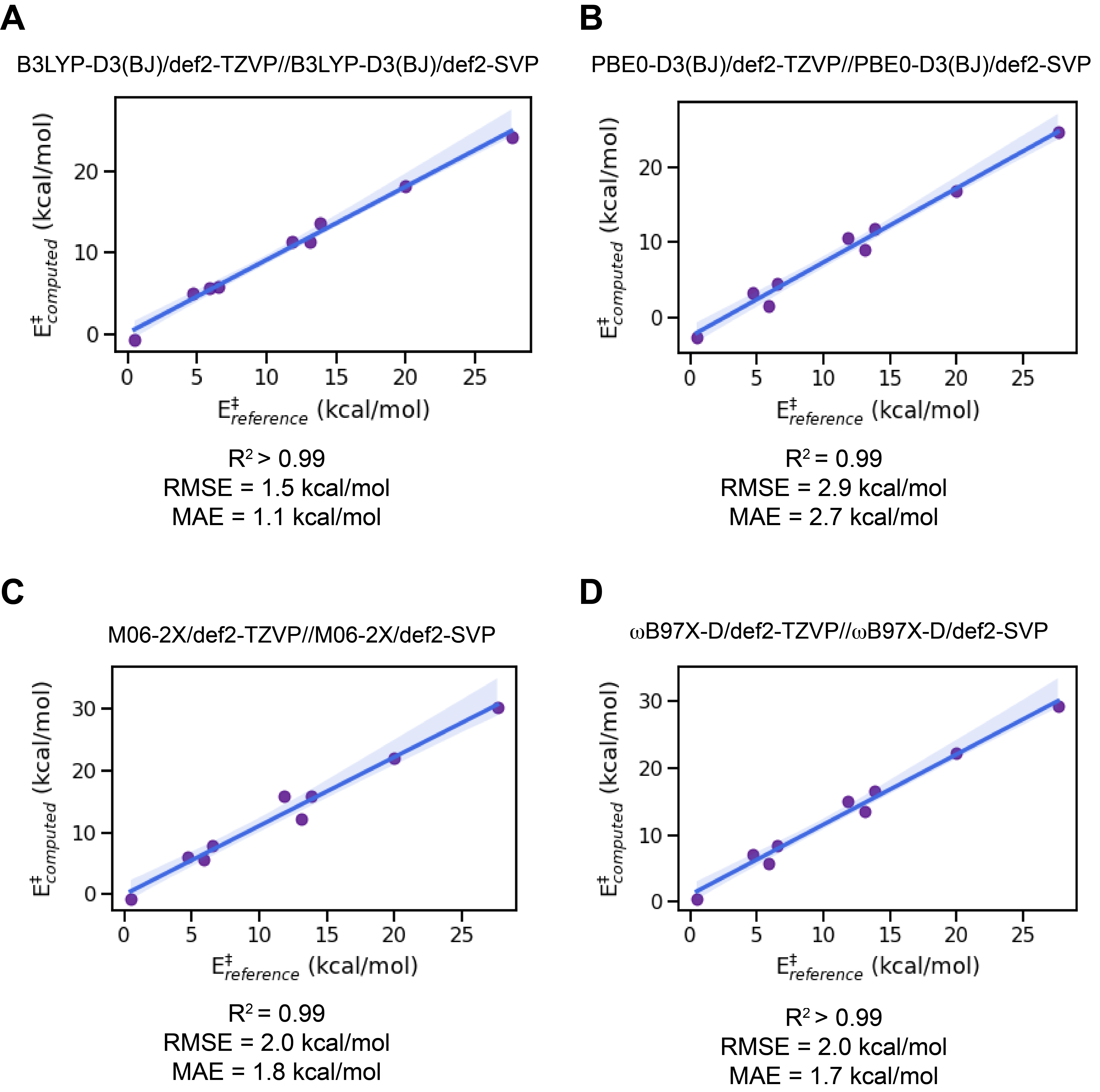}
\caption{Benchmarking functional + dispersion correction combinations against the (revised) BHPERI dataset.\cite{karton2015accurate} The translucent bands around the regression line correspond to the 95\% confidence interval determined through bootstrap resampling.}
\label{fig:benchmarking_gasphase}
\end{figure}

First, we selected the nine 1,3-dipolar cycloaddition reactions within the (revised) \cite{karton2015accurate} BHPERI dataset (see Figure \ref{fig:BHPERI}) \cite{goerigk2010general, ess2005activation} and compared the electronic energies to the Wn-F12 values from the literature. \cite{karton2012explicitly} All of the level of theories tested resulted in almost perfect correlations ($R^2 >=$ 0.99), but B3LYP-D3(BJ) clearly outperformed the other functionals in terms of mean-absolute-error (MAE; 1.1 kcal/mol vs > 1.8 kcal/mol) and root-mean-square-error (RMSE; 1.5 kcal/mol vs > 2.0 kcal/mol), see Figure \ref{fig:benchmarking_gasphase}. It should be noted that our findings here are also in line with the results presented in the very recent benchmarking study on pericyclic reactions by Bickelhaupt and co-workers: across the broader set of pericyclic reactions investigated, M06-2X was found to outperform B3LYP-D3(BJ), but specifically for the 1,3-dipolar cycloaddition barrier probed, B3LYP-D3(BJ) reproduced the CCSDT(Q)/CBS barrier within 1 kcal/mol, outperforming each of the other functionals by 1-3 kcal/mol. \cite{D2CP02234F}

\begin{figure}
\centering
\includegraphics[scale=0.44]{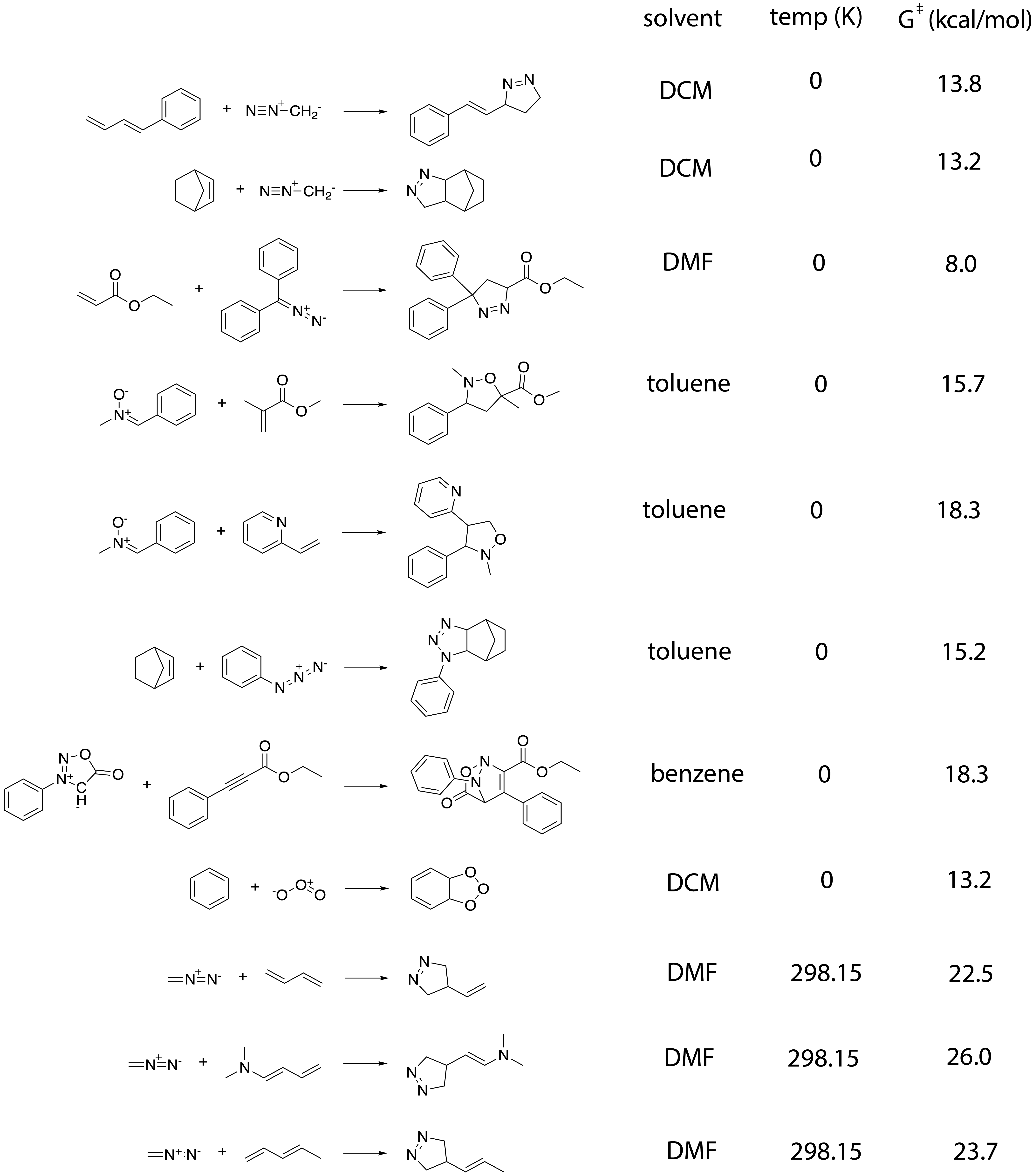}
\caption{Reference reaction data extracted from the literature.\cite{huisgen19841, geittner1978solvent, huisgen1963kinetics, huisgen19751} Temperatures of 0K indicate that the originally reported values are activation enthalpies $H^{\ddagger}_{0K}$ (obtained through extrapolation from measurements at different temperatures).}
\label{fig:benchmarking_solvent_data}
\end{figure}

Next, we curated an experimental benchmarking dataset based on rate constants found in a series of papers authored by Huisgen and co-workers on the kinetics (and mechanism) of 1,3-dipolar reactions. \cite{huisgen19841, geittner1978solvent, huisgen1963kinetics, huisgen19751} We transformed 11 reported experimental rate constants to Gibbs free energies of activation with the help of the Eyring equation assuming no recrossing (Figure \ref{fig:benchmarking_solvent_data}) and compared those to the corresponding values computed with our workflow (Figure \ref{fig:benchmarking_solvent}). \cite{eyring1935activated} Correlations between experimental and computed values were still excellent ($R^2 >$  0.92), but the best result was now obtained for $\omega$B97X-D ($R^2 =$  0.94). However, M06-2X and $\omega$B97X-D now yielded significant systematic errors (MAEs of 4.1-4.2 kcal/mol), whereas the experimental and computational barriers agreed much better for B3LYP-D3(BJ) and PBE0-D3(BJ) (MAEs of 2.7-2.8 kcal/mol). Putting all of this together, we decided to settle on the B3LYP-D3(BJ)/def2-TZVP//B3LYP-D3(BJ)/def2-SVP level of theory.

\subsection*{Generation of reaction profiles with autodE}

For every reaction SMILES generated with RDChiral, an autodE \cite{autodE} workflow to compute the associated reaction profile with the help of Gaussian16 \cite{frisch2016gaussian} was executed in a fully automated, high-throughput manner.  In first instance, the default settings of autodE were retained, and only the functionals and/or dispersion corrections were adjusted based on the benchmarking results (\emph{vide supra}). More information about this workflow can be found in the autodE documentation, and will not be discussed here at length.

We directly select the lowest energy conformer for reactant and product from GFN2-xTB \cite{bannwarth2019gfn2} optimization results for the generated conformer set, rather than from (single-point) DFT computations.\footnote{ We do so by setting the \texttt{ade.Config.hmethod\_conformers} keyword to False} 
While this approximation/simplification may be problematic when dealing with highly flexible compounds, it has a fairly limited effect on the results for our reactions, which involve mostly small and rigid compounds (cf. the excellent accuracy obtained during benchmarking and the tests performed on the azide reaction set; \emph{vide infra}). At the same time, this simplification resulted in a significant speed up of the reaction profile generation.

We also checked whether reactant and product complexes should be considered in the computed reaction profiles. By default, autodE does not compute complexes when thermal corrections (Gibbs free energies) are requested for the species along the profile, primarily because loose complexes with extremely low frequencies result in a high uncertainty for these quantities. 
To justify this choice, we compute reaction profiles for the (gas-phase) BHPERI dataset \cite{karton2015accurate} with both complexes and thermal corrections at room temperature. For each of the reactions for which the profile calculation terminated successfully, shallow minima for the complexes were found on the electronic energy surface, but thermal corrections erased these minima completely; the Gibbs free energy of the complexes consistently lay several kcal/mol above the isolated reactants (Table \ref{tbl:complexes}). Since electrostatic attraction -- which constitutes the bulk of the complexation interactions -- tends to diminish sharply when going from the gas-phase to a (polar) solvent, this result suggests that complex formation will not impact the reactivity for our cycloaddition reactions in a significant manner and can thus be safely neglected.

\subsection*{Ensuring stereochemical consistency between reactants, transition states, and products}
By default, autodE \cite{autodE} does not exchange stereochemical information between reactants and products in a reaction SMILES; it simply searches for the conformations of each species which have the lowest energy globally in an independent manner. The transition state (TS) therefore exclusively inherits the stereochemistry from one side. In the case of addition reactions, the product side stereochemistry is selected by default. Consequently, stereochemical compatibility between reactants, products and TSs is not guaranteed by default. As part of our workflow, a set of scripts was written to enforce such compatibility along the entire reaction profile.

\begin{figure}
\centering
\includegraphics[scale=0.6]{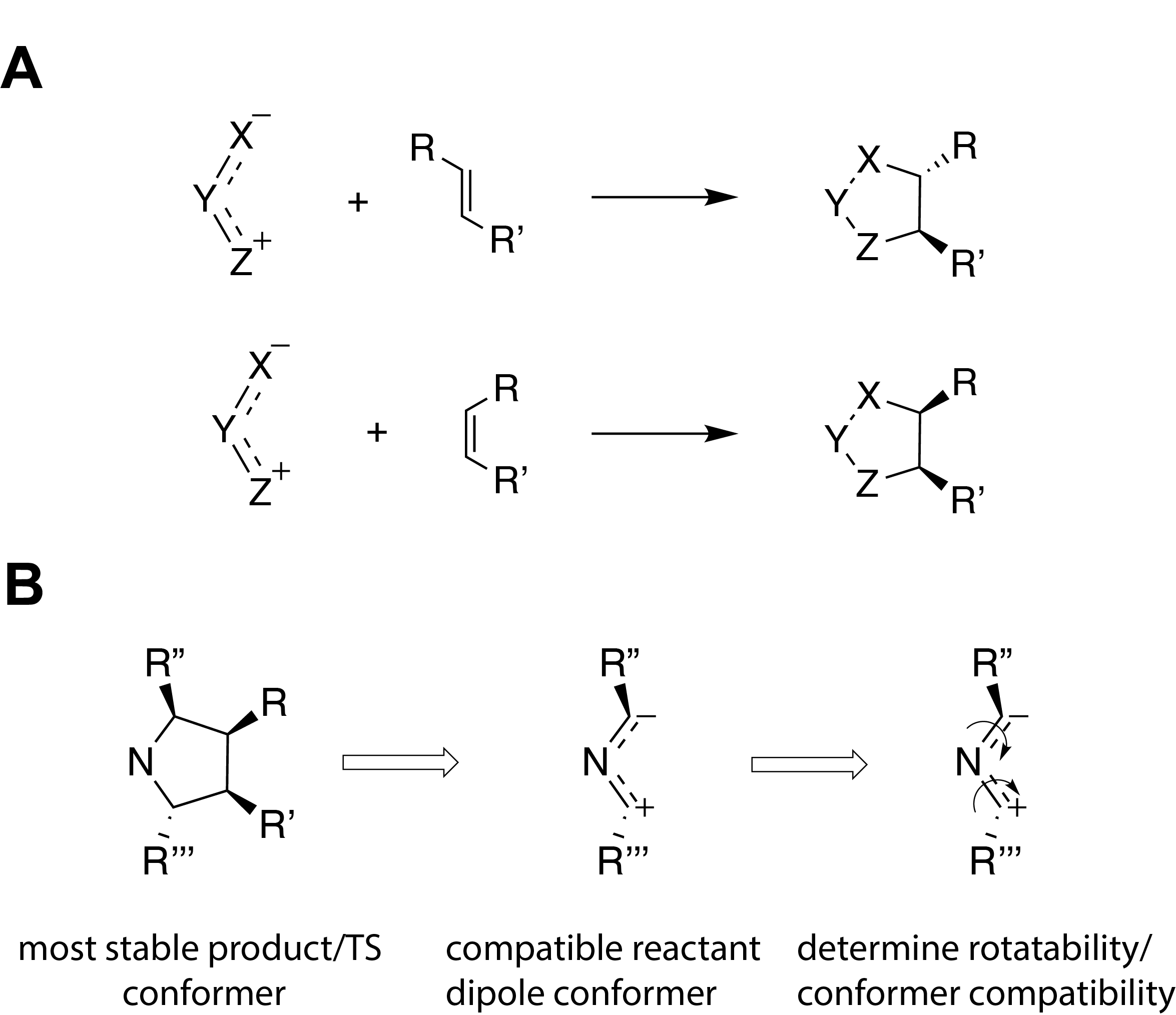}
\caption{(A) Illustration of stereo-retention around the unsaturated dipolarophile bond during a 1,3-dipolar cycloaddition reaction. (B) Schematic overview of the steps taken to ensure selection of a compatible dipole conformer: first, an approximate reactive dipole conformer is determined based on the selected (most stable) product/TS conformer geometry emerging from the reaction profile calculation); subsequently, the rotatability of the dipole bonds are assessed and the original reactant dipole conformer in the reaction profile is replaced by the (optimized) reactive one whenever rotation is determined to be too hindered.}
\label{fig:stereo_considerations}
\end{figure}

For the reactive sites, i.e., the atoms undergoing a change in bonding situation throughout the reaction, stemming from the dipolarophiles, it is possible to readily enforce stereo-compatibility by setting the stereotags in the SMILES representations outputted from RDChiral \cite{coley2019rdchiral} before the reaction profile computation is initiated, since doing so places constraints on the respective conformer search spaces in autodE. More specifically, we aimed to ensure that cis substituents in the reactant end up on the same side of the plane defined by the formed ring in the product (and \emph{vice versa} for the trans substituents), see Figure \ref{fig:stereo_considerations}A. 

To this end, we first verify that both centers undergoing addition are recognized as stereocenters in RDKit.\cite{landrum2013rdkit} Subsequently, the two potential product stereoisomers involving these centers are generated by setting the respective chiral tags, and guess structures are determined through a quick force field (MMFF94 \cite{halgren1996merck}) optimization. A similar optimization is performed for the reactant dipolarophile. Next, the dihedral angle for both potential products are compared to the corresponding dihedral angle in the reactants (0° or 180°). The product SMILES resulting in the lowest deviation in dihedral angle from the reactant geometry is then retained as the stereochemically correct product. 

For the reactive sites stemming from the dipoles, the situation is significantly more complex. First and foremost, it should be noted that for a significant number of scaffold types, there are no stereocenters at all (cf. the propargyl-allenyl ones), or these centers are so rigid that they are pre-set in practice (cf. the cyclic dipoles). For some allyl-type scaffolds however, particularly those involving two terminal C-centers, stereochemical considerations are relevant and compatibility needs to be enforced. Doing so in a similar manner as outlined for the dipolarophiles in the previous paragraph is not possible since the delocalization present in these dipoles causes the bonds connecting the individual centers to have bond orders in between 1 and 2, depending on the specific scaffold and the respective substituents. Consequently, either, both or neither of these bonds may be rotatable at room/body temperature.\cite{huisgen1967stereospecific} To complicate the situation even further, it is impossible to know \emph{a priori} which conformation around these partial double bonds will result in the lowest addition barrier to the specific dipolarophile considered. If no stereotags are specified for the dipole centers, autodE readily searches for the relative substituent orientation in the product that is most energetically favorable and the resulting arrangement will generally also be the lowest in the TS so that stereochemistry is generally retained in the second half of the reaction profile (\emph{vide supra}). From the product and TS geometries, the compatible reactant dipole conformer can be determined, but this can only be done after an initial version of the profile has already been generated.

As such, the following pragmatic approach was taken throughout this study. Initially, we did not fix any stereotag associated with the dipole in the reaction SMILES fed to autodE in either reactant or product. Once the full reaction profile was obtained, a script to correct potential stereochemical incompatibilities for the dipoles was executed (Figure \ref{fig:stereo_considerations}B). The coordinates of the subset of atoms in the TS geometry corresponding to the reactant dipoles are first extracted. Subsequently, the obtained geometry is optimized with GFN2-xTB.\cite{bannwarth2019gfn2} Next, a scan around the individual (partially double) dipole bonds is performed in 60 uniformly-spaced increments (GFN2-xTB level of theory; associated force constraint of 0.1 Hartree/Bohr$^2$) to obtain a crude estimate of the barrier for rotation. If the spread in energies (or the bias energy due to the constraining potential) along the profile exceeds 20 kcal/mol, then the bond is assumed to be more or less rigid, i.e., non-rotatable, under physiological conditions. Finally, a \emph{randomize and relax} (RR) conformer search is performed for each dipole within autodE.\cite{autodE} 1000 additional conformers (in addition to the one extracted from the TS geometry) are generated while constraining the dihedral angles around non-rotatable bonds. After initial pruning, conformers are optimized at GFN2-xTB level of theory and the most stable one is compared with the originally selected conformer in the reaction profile. If the RR conformer is lower in energy, it is retained as a new reference for the reaction profile: DFT optimization and single-point frequency calculations are performed in a follow-up step, the final species is included in the original output-folder for the reaction and updated activation and reaction energy values are computed. For dipoles with conformational restrictions (i.e., non-rotatable bonds), the lowest energy RR conformer is consistently selected as alternative reference for the reaction profile and the same procedure as described above is followed.

\section*{Data Records}

All data files produced as part of this study are accessible through Figshare (\url{https://figshare.com/articles/dataset/dipolar_cycloaddition_dataset/21707888}). \cite{Stuyver2022} Reaction IDs and SMILES, activation energies ($G^{\ddagger}$; in kcal/mol) and reaction energies ($G_r$; in kcal/mol) for each computed reaction profile are provided in CSV format (\texttt{full\_dataset.csv}). XYZ-files for each reactant (both the original and stereo-constrained versions), TS and product species as well as a CSV file containing computed electronic energies and thermal corrections are available in a compressed archive file, \texttt{full\_dataset\_profiles.tar.gz}. 

The files have been organized per reaction profile, identified through the reaction ID. Within each directory, reactant XYZ-files are of the form r\#\#\#\#\#.xyz, product XYZ-files are of the form p\#\#\#\#\#.xyz, and transition state XYZ-files are of the form TS\_\#\#\#\#\#.xyz. If the reactant dipole conformer had to be corrected to enforce stereochemical compatibility, the latter XYZ-files are included under to form of r\#\#\#\#\#\_alt.xyz. The energies for all of these species are summarized per directory in energies.csv.

Additionally, all the benchmarking data are made available in the \texttt{benchmarking\_data.tar.gz} directory (\emph{vide supra}).

\section*{Technical Validation}

The accuracy of the computed activation energies was assessed as part of the benchmarking study used to select the most appropriate DFT level of theory. As discussed in the Methods Section, errors on the (gas-phase) electronic energies were determined to be negligible for the selected functional (MAE  $\sim$ 1 kcal/mol), and even when thermal and solvent corrections are included, the errors relative to experimental activation energies remain sufficiently low to extract chemically meaningful trends from the data (MAE $\sim$ 2-3 kcal/mol). 

The established workflow is also quite robust, with a failure rate of 3.5\% during generation of the reaction SMILES from reactant combinations (mainly for the cylic dipoles) and 12.3\% during the ensuing autodE reaction profile computation and postprocessing. For one specific biologically inspired motif (the guanidinium moiety), we did encounter a much higher failure rate ($\sim$ 66\%). Our analysis suggests that these failures are caused by the disappearance of the cycloaddition reaction mode for the involved species, not an actual failure of our methodology (see below). 

To contextualize the numbers above, similar automated high-throughput reaction profile computation workflows achieve comparable or worse failure rates, e.g., a success rate of 85\% was achieved by Friederich \emph{et al.} in their study on dihydrogen activation of Vaska's complex \cite{friederich2020machine} and Von Rudorff \emph{et al.} were able to compute full reaction profiles for 25\% of the $E_2$/$S_N2$ reaction systems considered in their search space. \cite{von2020thousands} At the same time, even higher success rates have been reported as well, e.g., Jorner \emph{et al.} were able to obtain 98\% of the profiles in their study of nucleophilic aromatic substitution reactions. In the latter example, only reactions with a precedent in the experimental literature were considered, so "unrealistic" reactions were filtered out \emph{a priori}, which is not the case for most other workflows developed. \cite{jorner2021machine} 

\subsection*{Recovery of known bio-orthogonal click reactions from the presented workflow}

\begin{figure}
\centering
\includegraphics[scale=1]{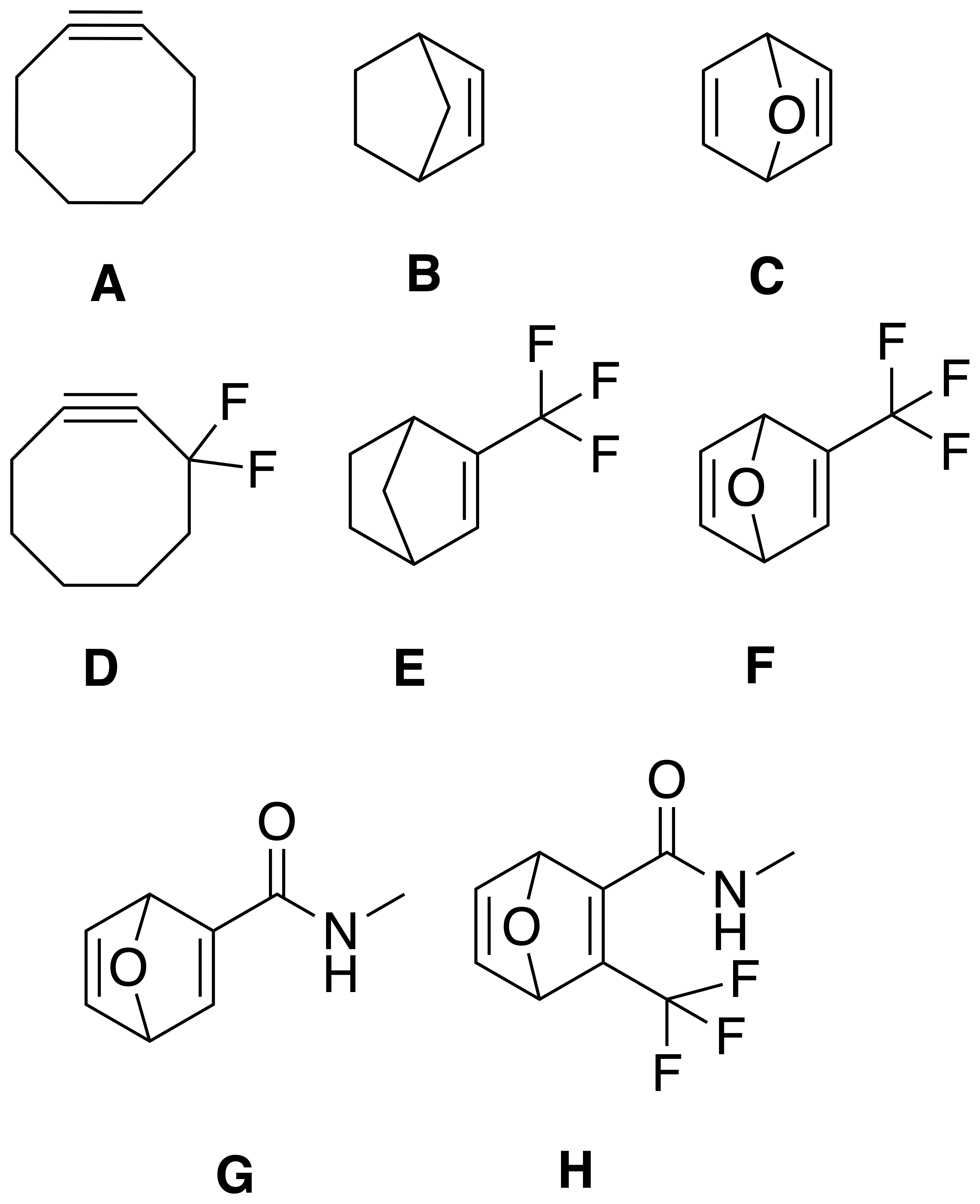}
\caption{The synthetic (strained) dipolarophiles included in the azide test reactions.}
\label{fig:azide_dipolarophiles}
\end{figure}

To further validate our workflow, activation energies for two prototypical azide-based 1,3-dipoles (methyl and acyl azide respectively) with a couple of popular strained dipolarophiles, such as cyclooctyne and oxo-norbornadiene, as well as all the selected biofragments were computed (see Figure \ref{fig:azide_dipolarophiles} for an overview of the selected synthetic dipolarophiles). 

In line with the results for the dataset as a whole, we find that our calculations fail for 3 out of 4 reactions with the tested guanidinium motif. This can be rationalized by considering the extremely unfavorable thermodynamics expected for these reactions: cycloaddition for these species requires the resonance in both the dipole and the guanidinium species to be broken, which imposes a significant delocalization penalty on any reaction mode involving these two species, and can potentially wipe out a barrier along the reaction pathway altogether (see Figure \ref{fig:azide_dipolarophiles}A). \cite{stuyver2020local} Support for this reasoning can be found in the fact that the reason for failure for each of these profiles was the inability by autodE to locate a stable molecule with the bonding pattern corresponding to the product, i.e., the expected product is not a minimum on the potential energy surface, and the observation that out of the -- in total -- 61 reaction profiles involving guanidinium which we were, in fact, able to compute successfully throughout the entire dataset, close to 95\% turned out to be endothermic, and over two-thirds are significantly so (endothermic by 20-65 kcal/mol). As an extra check, we attempted to compute reaction profiles for an alternative stepwise mechanism as well, but here the product tended to spontaneously decompose back into the reactants upon optimization.

\begin{figure}
\centering
\includegraphics[scale=0.39]{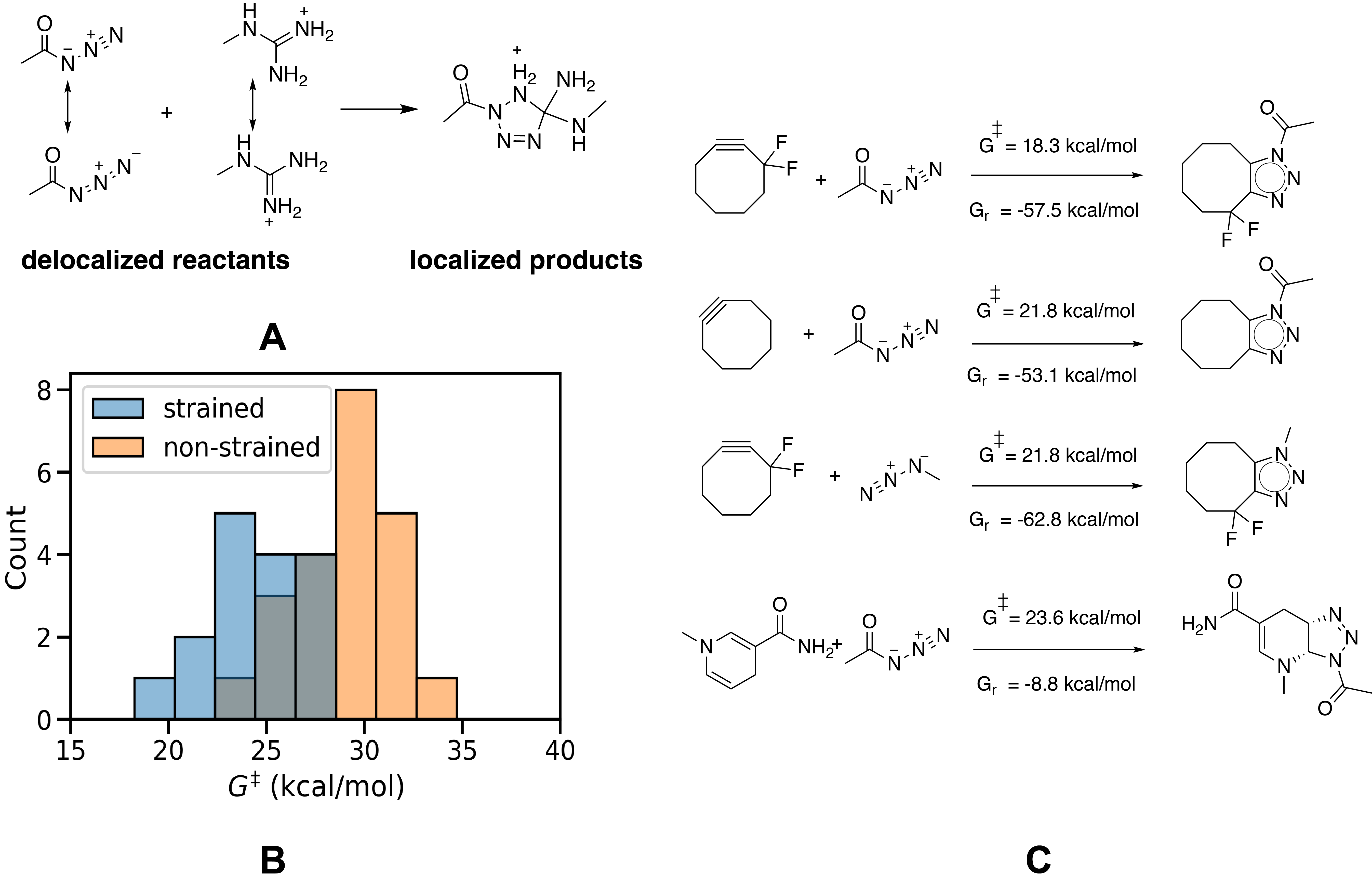}
\caption{(A) A schematic depiction of the resonance loss associated with the reaction between the guanidinium motif and acyl azide. (B) Histogram depicting the distribution of the activation energies ($G^{\ddagger}$) for both strained and non-strained dipolarophiles with methyl and acyl azide dipoles. (C) The three lowest reaction barriers computed for the strained dipolarophiles (top) and for the lowest non-strained dipolarophiles (bottom).}
\label{fig:histogram_azides}
\end{figure}

Focusing on the successfully computed reaction profiles, we find that most reactions between acyl/methyl azide and strained dipolarophiles exhibit a relatively low activation energy (between 21 and 25 kcal/mol under standard conditions), suggesting that most will readily proceed under physiological conditions (Figure \ref{fig:histogram_azides}). Furthermore, all of these reactions are highly exothermic, i.e., they are irreversible. The workflow readily identifies the reaction between (fluorinated) cyclooctyne and methyl/acyl azide as the most rapid transformations ($G^{\ddagger}$ = 18--22 kcal/mol; see the upper part of Figure \ref{fig:histogram_azides}B), underscoring the experimentally observed excellent click-potential of this popular dipole-dipolarophile combination. \cite{jewett2010cu}
On the other hand, reactions of the same azides with the biofragments tested tend to involve higher barriers (25 kcal/mol and up) on average, indicating that these will proceed at much slower/non-competitive rates than the ones involving synthetic/strained dipolarophiles. 

We found a single reaction involving biofragments with a relatively low reaction barrier ($G^{\ddagger}$ = 23.6 kcal/mol), which ought to be competitive with all but the most reactive ones involving strained dipolarophiles. This reaction involves a relatively exotic motif corresponding to (a fragment of) NADH/NADPH. While these compounds are intermediates of important biochemical processes \cite{ying2008nad+}, their concentrations tend to be relatively low; NADH can typically be found in mammalian cells at concentrations of approximately 10\textsuperscript{-4}--10\textsuperscript{-5}M, with most of it bound to proteins and hence only partially available for reaction. \cite{yang2007nutrient, blinova2005distribution} Consequently, even for synthetic reactions exhibiting similar activation energies, these native reactions will likely not disrupt the bio-orthogonal click function completely: unless the synthetic reaction is (significantly) slower and/or the synthetic dipolarophile is (much) less prevalent/available for the dipole partner, the latter will still dominate. 

The test case above underscores the subtleties -- and limits -- associated with bio-orthogonal click chemistry: even for tried and tested synthetic reactions, the difference in reactivity compared to the fastest native reactions tends to be rather subtle. Nevertheless, our computational approach enables us to retrieve the main reactivity trends for these [3 + 2] cycloadditions to identify promising/suitable bio-orthogonal click reactions.

\subsection*{Accuracy and reproducibility of the data}

Some additional tests to probe the accuracy and reproducibility of the data obtained through our workflow were performed on the azide reaction list described in the previous subsection. First, we considered the robustness and reproducibility of the cycloaddition reaction profiles generated with autodE in general. Since the conformer generation algorithm used in RDKit is stochastic in nature, two consecutive runs of autodE on the same reaction SMILES will not necessarily yield the exact same result since the sampled conformers can differ slightly. 

\begin{figure}
\centering
\includegraphics[scale=0.378]{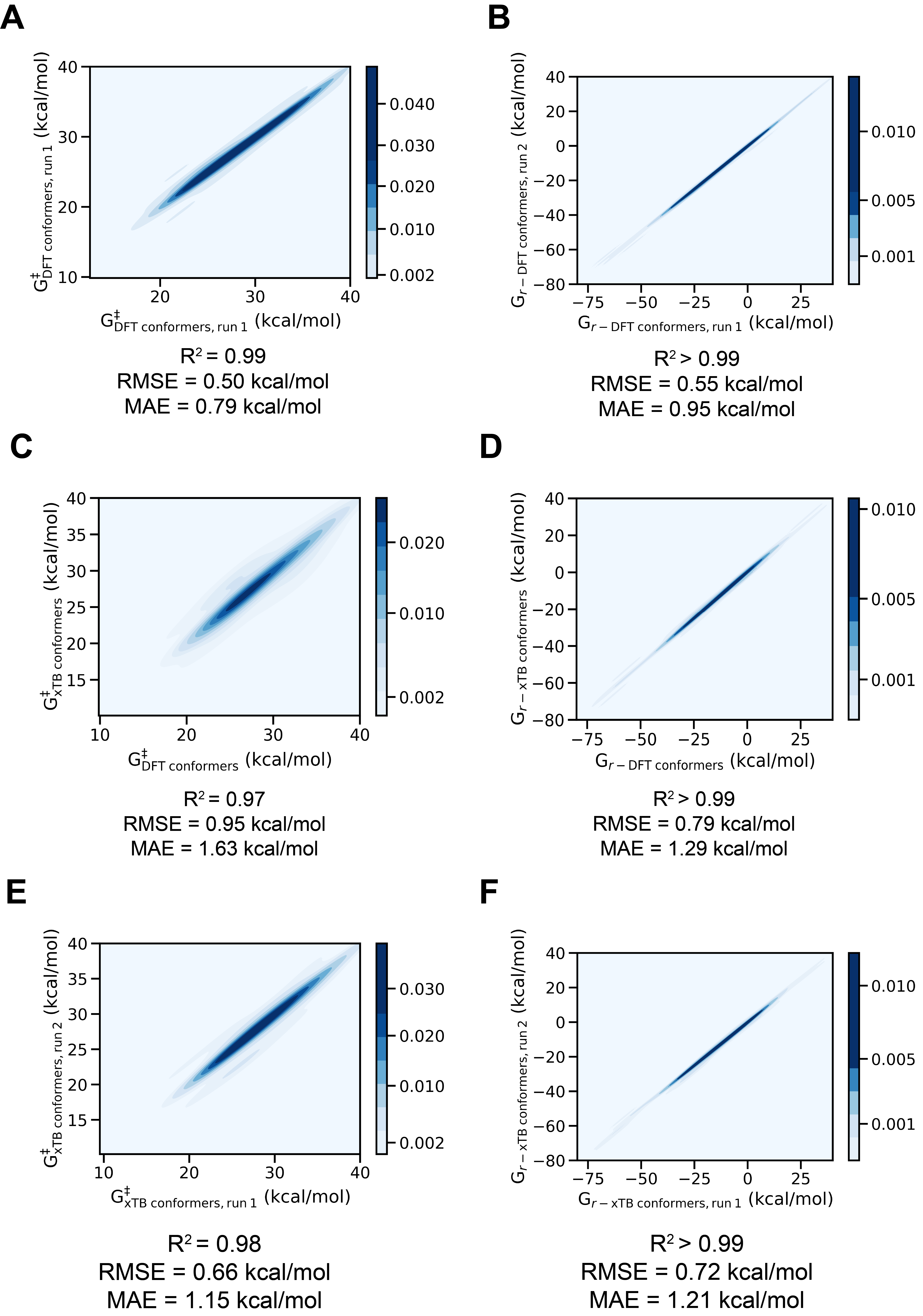}
\caption{(A) Correlation between the activation energies ($G^{\ddagger}$) and (B) the reaction energies ($G_{\text{r}}$) for two consecutive autodE runs with conformer selection at DFT level of theory for the azide test reactions. (C) Correlation between the activation energies and (D) the reaction energies for an autodE run with conformer selection at DFT level of theory and a consecutive run with conformer selection at xTB level of theory. (E) Correlation between activation energies and (F) the reaction energies for two consecutive autodE runs with conformer selection at xTB level of theory.}
\label{fig:azide_repeated_runs}
\end{figure}

To assess the magnitude of the resulting uncertainty, barrier heights and reaction energies for the azide reaction list were computed twice with the default autodE settings, i.e., conformers were generated with RDKit, after which the unique ones were optimized with DFT and the lowest-energy one was selected, and the resulting values were compared. As can be seen from Figure \ref{fig:azide_repeated_runs}A-B, activation and reaction energies for our test set are reproduced well (MAE $\sim$ 0.7 kcal/mol and RMSE $\sim$ 1.2 kcal/mol for the activation energies and MAE $\sim$ 0.8 kcal/mol and RMSE $\sim$ 1.3 kcal/mol for the reaction energies). 

Next, we assessed the effect of the approximation in conformer selection applied during the dataset generation. More specifically, all the barriers for the azide reaction list were again computed twice: in one run, conformers were selected based on GFN2-xTB conformer energies, in the second run, conformer selection was done in the default manner, i.e., based on DFT energies. As can be observed from Figure \ref{fig:azide_repeated_runs}C-D, the correlation between the barriers obtained in both runs is reasonably close to the reproducibility error at the most accurate conformer selection criterion (MAE $\sim$ 1.0 kcal/mol and RMSE $\sim$ 1.7 kcal/mol for the activation energies and MAE $\sim$ 0.8 kcal/mol and RMSE $\sim$ 1.3 kcal/mol for the reaction energies).

Finally, the reproducibility of the workflow with conformer selection at GFN2-xTB level of theory was checked. As can be observed from Figure \ref{fig:azide_repeated_runs}E-F, the reproducibility errors with and without the approximation are perfectly in line (MAE $\sim$ 0.7 kcal/mol and RMSE $\sim$ 1.1 kcal/mol for the activation energies and MAE $\sim$ 0.7 kcal/mol and RMSE $\sim$ 1.2 kcal/mol for the reaction energies).

Taking everything together, one can conclude that selecting conformers based on GFN2-xTB energies introduces only a small additional error relative to selection based on DFT energies, and thus we decided to consistently apply this approximation since this accelerates the workflow significantly, facilitating a broader exploration of the defined chemical space.

\section*{Usage Notes}

The code used to generate the reaction SMILES and automate the autodE workflow is available on GitHub. The repository contains several Python scripts and Jupyter Notebooks:
\begin{itemize}
    \item Notebooks to generate the respective search spaces and extract samples from them.
    \item A script to generate reaction SMILES with RDChiral and set the stereotags of the chiral centers associated with the dipolarophiles.
    \item A script and auxiliary modules to launch high-throughput reaction profile computation with autodE in a parallellized manner.
    \item A script to correct stereochemical incompatibilities in the selected reactant dipole conformers.
    \item A script and auxiliary modules to launch high-throughput DFT optimization and single-point fine-tuning of corrected dipole conformers.
\end{itemize}

\section*{Code availability}

The code described in the previous section is freely available in GitHub under the MIT license (\url{https://github.com/coleygroup/dipolar_cycloaddition_dataset}). Further details on how to use it is provided in the associated README.md file.

\section*{Acknowledgements}

T.S. and C.W.C. acknowledge the Machine Learning for Pharmaceutical Discovery and Synthesis Consortium for funding this work. K.J. acknowledges funding through an International Postdoc grant from the Swedish Research Council (No. 2020-00314). The authors acknowledge the MIT SuperCloud and Lincoln Laboratory Supercomputing Center for providing HPC resources that have contributed to the research results reported within this paper.

\section*{Author contributions statement}

T.S. and C.W.C. conceived the study, T.S. performed the computations and wrote the associated code, T.S. and K.J. developed the methodology and analyzed the data. All authors contributed to the writing of the manuscript.

\section*{Competing interests}

The authors declare no competing interests.

\newpage

\section*{Figures \& Tables}

\begin{table}
  \caption{Overview of the SMARTS templates used.}
  \label{tbl:SMARTS}
  \begin{tabular}{cc}
    \hline
    reactant system & SMARTS strings \\
    \hline
    \begin{tabular}{@{}c@{}} propargyl-allenyl dipole scaffolds \\ adding to double bonds \end{tabular} & \begin{tabular}{@{}c@{}} [*:1]$\#$[*+:2][*-:3].[C,N:4]=[C:5])>>[*:1]1=[*;+0:2][*;-0:3][C,N:4][C:5]1 \\ ([*:1]$\#$[*+:2][*-:3].[C,N:4]=[C:5])>>[*:1]1=[*;+0:2][*;-0:3][C:5][C,N:4]1  \end{tabular} \\
    \\
    \begin{tabular}{@{}c@{}} propargyl-allenyl dipole scaffolds \\ adding to triple bonds \end{tabular} & \begin{tabular}{@{}c@{}} ([*:1]$\#$[*+:2][*-:3].[C:4]$\#$[C:5])>>[*:1]1=[*;+0:2][*;-0:3][C:4]=[C:5]1 \\ ([*:1]$\#$[*+:2][*-:3].[C:4]$\#$[C:5])>>[*:1]1=[*;+0:2][*;-0:3][C:5]=[C:4]1  \end{tabular} \\
    \\
    \begin{tabular}{@{}c@{}} allyl dipole scaffolds \\ adding to double bonds \end{tabular} & \begin{tabular}{@{}c@{}} ([*:1]=[*+:2][*-:3].[C,N:4]=[C:5])>>[*:1]1[*;+0:2][*;-0:3][C,N:4][C:5]1 \\ ([*:1]=[*+:2][*-:3].[C,N:4]=[C:5])>>[*:1]1[*;+0:2][*;-0:3][C:5][C,N:4]1  \end{tabular} \\
    \\
    \begin{tabular}{@{}c@{}} allyl dipole scaffolds \\ adding to triple bonds \end{tabular} & \begin{tabular}{@{}c@{}} ([*:1]=[*+:2][*-:3].[C:4]$\#$[C:5])>>[*:1]1[*;+0:2][*;-0:3][C:4]=[C:5]1 \\ ([*:1]=[*+:2][*-:3].[C:4]$\#$[C:5])>>[*:1]1[*;+0:2][*;-0:3][C:5]=[C:4]1  \end{tabular} \\
    \\
    \begin{tabular}{@{}c@{}} cyclic dipole scaffolds \\ adding to double bonds \end{tabular} & \begin{tabular}{@{}c@{}} ([*:1]:[*+:2][*-:3].[C,N:4]=[C:5])>>[*:1]1[*;+0:2][*;-0:3][C,N:4][C:5]1 \\ ([*:1]:[*+:2][*-:3].[C,N:4]=[C:5])>>[*:1]1[*;+0:2][*;-0:3][C:5][C,N:4]1  \end{tabular} \\
    \\
    \begin{tabular}{@{}c@{}} cyclic dipole scaffolds \\ adding to triple bonds \end{tabular} & \begin{tabular}{@{}c@{}} ([*:1]:[*+:2][*-:3].[C:4]$\#$[C:5])>>[*:1]1[*;+0:2][*;-0:3][C:4]=[C:5]1 \\ ([*:1]:[*+:2][*-:3].[C:4]$\#$[C:5])>>[*:1]1[*;+0:2][*;-0:3][C:5]=[C:4]1  \end{tabular}
  \end{tabular}
\end{table}

\begin{table}
  \caption{Complexation energies for the cycloaddition reactions of the BHPERI dataset \cite{karton2015accurate}, where $E_{complex}$ stands for the electronic energy (+zero-point correction), and $G_{complex}$ stands for the Gibbs Free energy (at standard conditions, i.e., 298.15K and 1M).}
  \label{tbl:complexes}
  \begin{tabular}{ccc}
    \hline
    reaction SMILES & $E_{complex}$ (kcal/mol) & $G_{complex}$ (kcal/mol) \\
    \hline
    C=C.N$\#$[N+][O-]>>C1N=NOC1 & -3.2 & 1.9 \\
    C=C.N$\#$[N+]-[NH-]>>C1N=NNC1 & -3.6 & 3.3 \\
    C=C.N$\#$[N+]-[CH2-]>>C1N=NCC1 & -2.0 & 3.0 \\
    C=C.C$\#$[N+]-[O-]>>C1C=NOC1 & -2.9 & 2.6 \\
    C=C.C$\#$[N+][NH-]>>C1C=NNC1 & -3.2 & 4.0 \\
    C=C.C$\#$[N+][CH2-]>>C1C=NCC1 & -2.1 & 4.6 \\
    C=C.C=[NH+][O-]>>C1CNOC1 & -4.5 & 3.4 \\
    C=C.C=[NH+][NH-]>>C1CNNC1 & / & / \\
    C=C.C=[NH+][CH2-]>>C1CCNC1 & / & / \\
    \hline
  \end{tabular}
\end{table}

\end{document}